\documentclass[11pt]{article}
\usepackage{moriond,epsfig}

\begin{document}
\vspace*{4cm}

\title{\bf{ Theory Summary of the Electroweak Session for Moriond 2005}}

\author{R. D. Peccei\\}
\address{Department of Physics and Astronomy\\
University of California at Los Angeles\\
Los Angeles, California, 90095}
\vspace{1cm}

\maketitle\abstract{I broadly summarize the theoretical contributions in the Electroweak session of the 2005 Moriond meeting under four rubrics: i) neutrinos; ii) cosmology; iii) electroweak interactions; and iv) flavor physics.}

\section*{Introduction}

The theoretical talks in the Electroweak session  of the 2005 Moriond meeting were very lively and covered a broad spectrum of topics. Rather than presenting a detailed summary of what was discussed, I decided instead that it would be more useful to make some more general comments on the four "big topics" of the meeting: neutrinos; cosmology; electroweak interactions; and flavor physics. In all four areas considerable progress has occurred in the last few years and some interesting theoretical speculations have been put forth, which I would like to highlight here.

\section{Neutrinos: Windows to New Physics}

\subsection{What we know}

At Moriond, E. Lisi \cite{Lisi} reviewed thoroughly what we have learned in the last decade about neutrino masses and mixings. Physically, the neutrino mass eigenstates $\nu_i$ are not the same as the weak interaction eigenstates $\nu_{\alpha}$  associated with a given lepton flavor  $\ell_{\alpha}=\{e, \mu, \tau\}$, but are related by a unitary mixing matrix $U$:

\begin{equation}
 |\nu_{\alpha} > = \Sigma_iU^{\dagger}_{\alpha i}|\nu_i >
\end{equation}

In a three neutrino framework $U$ contains 3 angles and 3 phases and can be written as $ U= U_{\rm lept} V$, where $ U_{\rm lept}$  is the leptonic analog of the CKM matrix:

\begin{equation}
U_{\rm lept}=\left|\begin{array}{ccc}
c_{12}c_{13}&s_{12}c_{13}&s_{13}e^{-i\delta}\\
-s_{12}c_{23}-c_{12}s_{23}s_{13}e^{i\delta}&c_{12}c_{23}-s_{12}s_{23}s_{13}e^{i\delta}&s_{23}c_{13}\\
s_{12}s_{23}-c_{12}c_{23}s_{13}e^{i\delta}&-c_{12}s_{23}-s_{12}c_{23}s_{13}e^{i\delta}&c_{23}c_{13}
\end{array}\right|
\end{equation}
and
\begin{equation}
V=\left |\begin{array}{ccc}
e^{i\alpha_1/2} &0&0\\
0&e^{i\alpha_2/2} &0\\
0&0&1
\end{array}\right|.
\end{equation}

 Atmospheric neutrino oscillation experiments are consistent with maximal mixing, which implies that $ s_{23} \simeq c_{23 }\simeq 1/\sqrt{2}$. Lisi's fit gives
\begin{equation} 
      s_{23}^2 = 0.45^{+0.18}_{ -0.11}.  
\end{equation}
The best fit of all solar neutrino oscillation data is the Large Mixing Angle MSW solution, where one has, approximately, $s_{12}\simeq 1/2 ;~ c_{12} \simeq \sqrt{3}/2 $. More precisely, Lisi finds
\begin{equation}
    s_{12}^2 = 0.29^{ +0.05}_{ -0.04}. 
\end{equation}
Strong bounds exist on the oscillations of reactor neutrinos, coming from the CHOOZ \cite{CHOOZ} and Palo Verde \cite{PV} experiments, which are confirmed by the full three neutrino analysis presented by Lisi, who finds
\begin{equation}
      s_{13}^2 < 0.035~~  [2 \sigma ~\rm{level}].  
\end{equation}

In the three neutrino framework, the neutrino oscillation results also identify the 2-3 mass difference with the mass difference measured in atmospheric oscillations and the 1-2 mass difference with that associated with solar neutrino oscillations. One finds
\begin{eqnarray}
 | \Delta m_{23}^2|& =& \Delta m^2_{\rm{atmos}} \simeq 2.4 \times 10^{-3}~\rm{ eV}^2\\ 
 | \Delta m_{12}^2| &=& \Delta m^2_{\rm{solar}}  \simeq 8 \times 10^{-5}~\rm{ eV}^2.
\end{eqnarray}
The result of the LSND experiment \cite{LSND} on $\bar{\nu}_{\mu} \to \bar{\nu}_e$ oscillations, which corresponds to a mass difference squared $ \Delta m^2\sim  1~\rm{ eV}^2$ and a mixing angle   $sin^2 \theta \sim~ 10^{-3}$ , cannot be reconciled with the above results in a three neutrino framework. If  the LSND result is true it would require introducing physics beyond the Standard Model, like sterile neutrinos, \cite{sterile} or perhaps even violations of CPT. \cite{CPT} We all await with great interest the results of the Mini BooNe experiment at Fermilab (reviewed here by McGregor \cite{McG}).

The data on neutrino oscillations gives no information on the Majorana phases $\alpha_1$ and $\alpha_2$ and the present data also does not determine the other CP-violating phase $\delta$. This latter phase is, in principle, measurable by comparing neutrino and antineutrino oscillations. However, to see an effect it is necessary that the mixing angle $\theta_{13} \neq$ 0. Note also that present data does not fix the neutrino spectrum, since what is measured in oscillation experiments are mass differences squared, $\Delta m^2$ . However, because $|\Delta m^2_{23}|= \Delta m^2_{\rm{atmos}}$ is much greater than $| \Delta m^2_{12}|= \Delta m^2_{\rm solar}$, it is reasonable to imagine a hierarchical neutrino spectrum with $ m_3 >> m_2\simeq m_1$ (or an inverted hierarchy, with $ m_2 \simeq m_1 >>m_3$ ).

In addition to the neutrino oscillation results described above, three "direct " mass measurements all give bounds  for  neutrino masses in the eV range. In tritium beta-decay one measures the effective mass $(m_{\beta})_{{\rm eff}} =[\Sigma_i | U_{ei} |^2 m^2_{{\nu}_i }]^{1/2}$. In neutrinoless double beta-decay one measures $  <M_{ee}>  = | \Sigma_i  (U_{ei} )^2 m_{{\nu}_i} |$ , while cosmological considerations \cite{Pastor} provide a bound on $ \Sigma = \Sigma_i  m_{{\nu}_i}$. Obviously, it would be wonderful if instead of just having bounds on these quantities we could eventually get an actual measurement!

 \subsection{What we would like to know}

We will learn much more about neutrinos in the future and, as Boris Kayser \cite {Kayser} stressed at Moriond, there are a set of critical questions which we would like to have answers for. These are:

  i.           {\it Are there more than 3 flavors of neutrinos? }\newline  
\noindent We know from LEP that the number of active neutrinos (neutrinos that have $SU(2)\times U(1)$ quantum numbers), $N_{\nu}= 2.984 \pm 0.008$  is  consistent with having only three flavors. However, are there sterile neutrinos? For that we must really check whether the LSND result is correct or not through the Mini BooNe experiment.

ii.            {\it Do neutrinoless double-beta decay processes exist?} \newline
\noindent The level of accuracy which is interesting and may be achievable in future experiments is $<M_{ee}>\simeq $ 0.1 eV- a level discussed at Moriond by  Capelli \cite {Capelli} for the forthcoming Cuore experiment. Seeing a signal for neutrinoless double beta-decay will tell us that neutrinos are self-conjugate Majorana particles and that indeed Lepton Number is not a symmetry of nature. This result will provide an experimental basis for the celebrated seesaw mechanism. \cite{seesaw} 
The most general neutrino mass term 
\begin{equation}
{\cal{L}}=-\frac{1}{2}(\bar{\nu}^c_L~ \bar{\nu}_R)\left|\begin{array}{cc}
m_T &m_D^T\\m_D&m_S \end{array}\right| \large{(}\begin{array}{c} \nu_L\\ \nu^c_R\end{array} \large{)}
\end{equation}
  does not conserve Lepton Number if $m_T, m_S \neq 0 $.  If $m_T<< m_D << m_S$ , the mass matrix will have a set of large eigenvalues connected with $m_S$, and a set of small eigenvalues connected with the matrix $M_{\nu}= m^T_Dm_S^{-1}m_D$.  

    iii.            {\it What is the value of $s_{13}$}?\newline
\noindent  A likely goal here is to be sensitive to  $s_{13}^ 2 \simeq 0.01$, both through reactor disappearance experiments and in accelerator appearance experiments. With the latter experiments, as Kayser emphasized, by measuring differences between neutrinos and antineutrinos one may get information on whether the neutrino spectrum is normal or inverted. In this context, Tanimoto \cite {Tan} discussed how $ s_{13}=0$ can be obtained by imposing a discrete $Z_2$ symmetry on the theory. He showed that if one asks, in addition, for maximal mixing, $s_{23} = 1/\sqrt{2}$, this leads to a neutrino mass matrix of the form
\begin{equation}
M_{\nu}=\left |\begin{array}{ccc}
X&A&A\\
A&B&C\\
A&C&B
\end{array}\right |.
\end{equation}
It is interesting to ask whether one can connect the departures of $s_{23}$ from $1/\sqrt{2}$ with that of $s_{13}$ from 0, as a way to try to predict this latter angle. A quaternionic model where this happens was discussed at this meeting by Frigerio. \cite{Frigerio}

   iv. {\it Is there CP-violation in the neutrino sector?} \newline
 \noindent          Eventually, one wants to observe directly CP-violation in the neutrino sector, through the measurement of the CP-violating phase $\delta$.
To achieve this, one needs a  2 MW proton driver to get enough neutrino flux. \cite{Kayser} However, this may not suffice since the signal is proportional to $s_{13}$,     
and one probably needs to have  $s_{13}^2 > 0.01$ to actually observe a signal.
Note that  CP-violation in the neutrino sector is necessary for Leptogenesis- probably the most appealing scenario for generating the matter- antimatter asymmetry 
in the Universe- although the CP phase that enters in Leptogenesis is not necessarily directly related to $\delta$.

\subsection{Cosmological issues}

The reason I am so sanguine about having thermal Leptogenesis be the origin of the matter-antimatter asymmetry in the Universe is that this mechanisms gets the right value for the ratio of the baryon to photon densities $\eta=n_B/n_{\gamma}$  measured by WMAP \cite{WMAP} and in big bang nucleosynthesis (BBN) \cite{BBN} for neutrinos which have {\bf {precisely}} the properties we observe. \cite{BPY} Indeed, so as to prevent a wash-out of the asymmetry generated by Leptogenesis \cite{BDP} one needs that $ m_{{\nu}_i} < 0.12$ eV.

However, as Turzynski \cite{Tur} discussed, there are conflicts between thermal Leptogenesis and expectations in supersymmetric (SUSY) theories. In particular, Leptogenesis requires that the mass of the lightest superheavy neutrino $ M_1> 2 \times 10^9 $ GeV and, in SUSY theories, this leads to an overproduction of gravitinos. Gravitino decays can alter the predictions of BBN and, typically, this constrains the reheating temperature of the Universe to $T_R<10^7$ GeV $<<M_1$. \cite{Moroi}
There are solutions to the gravitino problem, \cite{BPY} but these in general alter the  "normal" SUSY expectations [eg. $m_{3/2} >100$ TeV; gravitinos are the LSP; etc].

A different kind of "cosmological" tension was noted by Pastor \cite{Pastor} and Lisi. \cite{Lisi} They observed that  the strictest bound on the sum of neutrino masses $\Sigma <0.47$ eV, obtained  by asking that the density of free streaming massive neutrinos be small enough so as not to alter  the power spectrum at small scales, and the recent claim of a non zero value for $<M_{ee}>$  by Klapdor {\it {et al}} \cite{K} in the  Heidelberg-Moscow $^{78}$Ge experiment are mildly incompatible. \cite{Fogli} Obviously, matters will become very interesting if one can push both results near the 0.1 eV level.

\section{Cosmology: questions and opportunities}

      \subsection{Dark energy and particle physics}

We know from recent cosmological observations that the energy density of the Universe is dominated by a dark energy component, whose negative pressure causes the Universe's expansion to accelerate. One finds, approximately, that: \cite{WMAP} $\Omega_{DE}\simeq 0.73; \Omega_{DM}\simeq 0.23; \Omega_B\simeq 0.04$. Not only does the dark energy dominate, but the matter content is predominantly made up of non- luminous (dark) matter, with baryons accounting for less than 5\% of the energy density in the Universe today. Observations also bound the ratio $w=p_{DE}/\rho_{DE}$ of pressure to energy density in the Universe now to the range $-1.15<w<-0.8$, with the value $w=-1$ corresponding to a cosmological constant.

 The theoretical implications of these results were discussed by Frampton at Moriond. \cite{Frampton} Frampton spent considerable time discussing the future of the Universe and the physics associated with having $ w~<-1$. I consider this "Big Rip" scenario as unphysical, since it is connected to a negative energy density. Indeed, as Frampton showed, by means of a Lorentz  transformation, if $w ~<-1$, one can change the sign of the energy density:      $\rho^{\prime}= \rho(1 + \beta^2w)$.

Rather than focusing on the future, I prefer to focus on past. If the ratio $w$ is constant, it is easy to see from the energy conservation equation
\begin{equation}       
\frac{\partial \rho_{DE}}{\partial t} =-3H(\rho_{DE}+ p_{DE})= -3H\rho_{DE} (w +1)
\end{equation}
 that $\rho_{DE} \sim    R ^{- 3(1+ w)}$. Since $\rho_{\rm{Matter}}\sim R^{-3}$, if $w$ is a constant - like it would be in the case of a cosmological constant- in earlier epochs dark energy was subdominant. Obviously, a  
crucial cosmological question which needs answering is whether $w$ evolves or not. That is, is $w=w(T)$?

 From the point of view of  particle physics, the principal question to ask is what is the nature of the dark energy. For instance, if the dark energy were simply due to a cosmological constant, then $\rho_{DE}$  is a pure vacuum energy density and one has precisely $w=-1 $, since
\begin{equation}
      \rho_{DE} = -p_{DE} = E_o^4.
\end{equation}
 Experimentally, one finds that the energy scale $E_0$ is very small [$E_o \simeq 2 \times 10^{-3}$ eV], since the Hubble constant now corresponds to a tiny scale, of order $10^{-33}$ eV. Such a "small" vacuum energy is  very difficult  to contemplate in particle physics. For instance, vacuum energies associated with gluon or quark condensates in QCD  have typical scales of order: $E_o^{QCD} \sim \Lambda_{QCD} \sim $ 1 GeV.

In the Electroweak session of Moriond this year we had no discussion on possible particle physics approaches to the dark energy problem. However, both Frampton and Tyniakov \cite{Tinya} discussed models where dark energy results from possible modifications of gravity. Frampton \cite{Frampton} discussed the implication of higher dimensional modifications of gravity due to Dvali {\it{et al}} \cite{Dvali} involving a new fundamental length $L= M^2/M^{*3}$, related to ratios of the Planck constant in $d=4$ and $d>4$ dimensions. In these theories, dark energy is essentially mimicked by terms coming from the $d>4$ theory. Tinyakov, on the other hand, presented a model leading to a massive graviton. However, because Lorentz invariance is broken in the theory he considered, the potential for this model reduces to the standard gravitational potential 
\begin{equation}
  V=\frac{1}{r} + m_2^2 r F(\mu) \to \frac{1}{r}, 
\end{equation}
 because the function $F(\mu)$ can be chosen to vanish- a freedom allowed by the Lorentz breaking. The massive graviton, is  strongly bounded observationally and its mass must be tiny [$m_2< 10^{-20}$ eV]. Effectively, the massive graviton acts as cold dark matter, while dark energy  in the model is, essentially, a cosmological constant. Although the models discussed at Moriond  are theoretically interesting, frankly they raise more questions than they answer!

       \subsection{Conventional and unconventional dark matter}

At Moriond there was more discussion on the nature of dark matter (DM) itself. This DM is really normal matter, with $p=0$, but which, however, is non-luminous. Most of the discussion revolved around possible observational signals for DM in non-accelerator experiments, as well as a variety of theoretical considerations. As is well known, one can look for either indirect signals of DM, resulting from DM annihilations in the galaxy, or direct signals from DM interactions in a detector. I will not discuss here any of the details of the purported indirect or direct experimental signals for DM, as this is not my task, but will limit myself to some broad theoretical observations

There are really not many bonafide particle physics ideas for what dark matter may be. In fact, in my view, there are only two candidates  which have some true theoretical motivation. Namely,

        i.            Axions, which to be a  viable DM candidate need to be associated with a $U(1)_{PQ}$ \cite{PQ} symmetry breaking scale of order $f_a \sim 10^{12}$ GeV or, equivalently, an axion mass of order $ m_a=10^{-6}$ eV.

      ii.            A stable neutral supersymmetric particle , the SUSY LSP, which could be a neutralino, a gravitino, or even possibly a sneutrino.

Most of the efforts in the field, perhaps naturally, has been expended in analyzing SUSY DM in  the simplest supersymmetric extension of the standard model, the so called MSSM framework. One should  be cognizant, however, that this may be too naive an assumption. For example, if thermal leptogenesis is the origin of the matter-antimatter asymmetry in the Universe, the gravitino problem makes it very unlikely that the neutralino is the LSP. If it is, the SUSY spectrum may well be quite different from what is assumed in the MSSM. A case in point is the model that Profumo \cite{Profumo} discussed at the meeting  where one has a nearly degenerate chargino/neutralino pair with $\Delta [m_{\chi^+} - m_{\chi^o}]/m_{\chi^o} \sim 10^{-3}$, very different from the expectations of the MSSM. What is clear, however, as discussed by Nezri \cite{Nezri} is that the LHC will have enormous bearing on the question of supersymmetric dark matter.

Regarding some of the possible hints for DM in non accelerator experiments discussed at Moriond, it is important that the theoretical analysis of the claimed signals have some cross checks. In this respect,  the detailed analysis by Boehm \cite{Boehm} of the 511 keV signal from INTEGRAL is perhaps not so well grounded theoretically. Although using a better motivated density profile clearly strengthens the claim, the fact remains that the motivation for a DM candidate with $m_{DM} < 100$ MeV is rather questionable. In this respect, the analysis of de Boer \cite{dB} of a possible DM signal in the diffuse photon background measured by EGRET is more promising. The location of the purported EGRET signal
 determines kinematically the mass of the DM particle responsible for the photon excess to be in the neighborhood of  $   M\sim 50-100$ GeV. Once this is known, there is a clear prediction for the cross section expected in direct searches of this particle: $\sigma \sim 2 \times 10^{-43}~\rm{cm}^2$. This number is only slightly below the present limit from CDMS \cite{CDMS} for a DM particle of this mass: $\sigma < 7 \times 10^{-43}~\rm{cm}^2$, so eventually this signal can be confirmed or ruled out.

\subsection{Matter-antimatter asymmetry and dark matter}

There is another important ratio in cosmology that one would like to understand, the ratio of the DM to baryon density in the Universe, $\Omega_{DM}/\Omega_B \simeq 6$. In supersymmetric models, in general, the mechanism which generates the
matter-antimatter asymmetry is not directly connected to SUSY dark matter. Even if both were due to supersymmetric phenomena [e.g.  the asymmetry arises as the result of Affleck-Dine baryogenesis \cite{AD} and neutralinos are the DM] the physical scales associated with these phenomena are unrelated [In the example considered, the baryogenesis scale is associated with the lifting of some flat direction, while the neutralino mass is related to the scale of Electroweak symmetry breaking]. Thus, in these models, there is  no clear expectation for the ratio $\Omega_{DM}/\Omega_B $.

The situation is different in the case where Leptogenesis is the source of the matter-antimatter asymmetry and axions are the DM. In
Leptogenesis $\Omega_B$ is proportional to the mass of the lightest right-handed neutrino:  $\Omega_B \sim M_1$. If  axions are the dark matter, on the other hand, $\Omega_{DM} \sim f_a$ , where $f_a$ is the scale of PQ symmetry breaking. \cite{Turner} However, it is very natural to link $ M_1$ and $f_a$. Indeed, one can imagine the right-handed neutrino mass $M_1$ resulting from spontaneous symmetry breaking: $M_1\sim <\sigma>$. Instead of carrying Lepton Number, the $SU(2) \times U(1)$ singlet field $\sigma$ can carry a PQ charge, in which case $f_a = <\sigma>$. \cite{LPY} Thus both phenomena occur at same scale and $f_a$ drops out in the ratio  $\Omega_{DM}/\Omega_B $.

   \section{Electroweak theory}

    \subsection{Refinements}

        Precision electroweak data is in excellent agreement with the   $ SU(2) \times U(1)$ model . A global fit of all this data strongly points to a light Higgs boson, with the result of the fit giving: \cite{EW}
\begin{equation}
 M_H= 126^{+73}_{ -48 }\rm{GeV}.
\end{equation}
Final LEP numbers are in preparation and updated numbers are coming from the Tevatron. These results should help refine the analysis even further, but will probably not cause major changes to the present fit. At the moment, only the NuTeV \cite{NuTeV} result on the weak mixing angle (which can be translated into an effective measurement of the W- mass) is glaringly discrepant. NuTeV finds
$M_W= 80.136\pm 0.084$, to be contrasted with the LEP average $ M_W=80.412 \pm 0.042$. It is possible that there are larger theory errors than assumed, associated with heavy quark structure functions and other QCD uncertainties incurred in trying to extract a value of $sin^2\theta_W$ from the measured NuTeV data, but this has not convincingly been demonstrated.

The next logical step for testing the electroweak theory is actually discovering the Higgs boson, along with reducing further (where possible) the errors on measured parameters, like the top mass. With the Higgs mass in hand, the prototypical test is to confront the experimental value for, say, the W-mass $[M_W]_{\rm{exp}}$ with the now totally predictable (in the Standard Model) theoretical value $ [M_W]_{\rm{theo}} =M_W(m_t, M_H)$. There is a possibility that the  Higgs boson will be discovered at the Tevatron, but as Bernardi remarked \cite{Bernardi} to do this both CDF and D0 will need to integrate as much luminosity as possible [$ \int L dt \simeq 4-8 ~{\rm fb}^{-1}$]. However, the  LHC is really the machine where it will become clear what the nature of Electroweak symmetry breaking is. Is it simply the result of a single Higgs VEV, leading to one Higgs boson, or  is it something much more complicated?

     \subsection{Non-standard ideas}

Theorists, for a variety of reasons, believe that just a single Higgs VEV, with its associated Higgs boson, is unsatisfactory. In a theory with a physical cutoff $\Lambda$, the Higgs mass squared gets a quadratic shift from radiative corrections, $ \delta M_H^2 \sim \alpha \Lambda^2$, which is typically much bigger than the original mass squared. This is the 
Hierarchy problem. As is well known, it can be avoided either because some underlying symmetry changes the above formula (for example, if there is a low energy supersymmetry one has  $ \delta M_H^2 \sim \alpha M_H^2 \ln \Lambda/M_H$),  or the cutoff is very near (for example, in Technicolor theories, $\Lambda \sim M_H$). This latter option is disfavored by the electroweak data which are fit very well with no structure, except for a light Higgs.

Theorists continue to explore other ideas besides the  above classic alternatives, and Moriond was no exception, with the new game in town being theories in $d>4$ dimension. Here I will make some comments on three topics which were discussed at the meeting (which are really much more ideas than full fledged theories!), in increasing order of wildness:

        i.            Attempts to reconcile the Higgs mass cutoff $\Lambda\sim 1$ TeV, with the scale $\Lambda_{\rm{eff}} \sim 10$ TeV  which emerges when one bounds, using experimental data, the scale associated with irrelevant operators in the Standard Model 
\begin{equation}
{\cal{L}_{\rm eff}}  = {\cal{L}_{\rm SM}}   + \Sigma_i \frac{O_i^D}{\Lambda_{\rm eff}^{ D-4}}.
\end{equation}

      ii.            Revival of Technicolor theories as $d>4$ theories. 

    iii.            Landscape picture  and split SUSY theories.

In supersymmetric theories the "little "hierarchy $\Lambda_{\rm eff} \sim 10\Lambda$ is naturally satisfied by the usual loop expansion  relation $\Lambda_{\rm eff} =4\pi \Lambda$.  Biggio \cite{Biggio} described how  the same effect can  occur in $d>4$ theories, when the extra dimensions are compactified in orbifolds. In these theories the Higgs fields are part of the higher dimensional gauge field in the compact directions, $A^M=\{A^{\mu}, A^a=H_a\}$, and the Higgs mass is partially protected by the $d>4$ gauge symmetry. However, there are subtleties connected with what happens at specified points in the orbifold and this works only for certain $d>4$ theories. For, example, it does not work for d=6 because of the presence of effective Higgs mass terms coming from tadpole terms like $\epsilon^{ab}H_aH_b$.

Hidalgo \cite{Hidalgo} talked about another possible solution to the little hierarchy problem- Little Higgs Models. In these models the Higgs mass is partially  protected because it is a pseudo-Goldstone boson of a global symmetry which holds at the scale $\Lambda_{\rm eff }$ and which is spontaneously broken at a scale $\Lambda$. Because of this extended structure, these models have new states with masses of $O(\Lambda)$. These theories work technically, but introduce an enormous amount of superstructure which make them not very believable.

More appealing, but still very speculative, are the $d>4$ Technicolor models discussed by Grojean. \cite{Grojean} These models, by construction, have {\bf{no Higgs bosons}}. However, what physically replaces the single Higgs boson of the SM are towers of Kaluza Klein states. These states effectively  serve to give the required cancellations in the Electroweak amplitudes, effected by the Higgs boson in the SM, which preserve unitarity.
The models discussed by Grojean are technically complicated, as one must build up "by hand" much of the structure present in the SM. For example, the hidden $O(3)$ symmetry of the SM Higgs potential which is responsible for setting $\rho =1$ here arises from having a $d>4$ space which is highly weighted towards a brane with an SU(2) symmetry. However, once this is assumed, then the models naturally produces values for the S and T parameters which agree with experiment. Nevertheless, much work needs to be done to make realistic models of this type!

Some of today's most speculative ideas were  wonderfully reviewed by Dudas. \cite{Dudas} Of all the topics he discussed, I will talk  briefly here only about one such idea- the SUSY landscape. The idea of having a very large number of possible quantized vacuum states in string theory - the landscape-  is naturally related to the enormous hierarchy associated with the cosmological constant : 
\begin{equation}
(\frac{M_P}{ E_o})^4~\sim 10^{ 125}.
\end{equation}
From this point of view, our Universe, with its very small cosmological constant of $O(E_o)^4$, emerges from this plethora of states through the anthropic principle.

Having sinned once (and in a spectacular way!) by appealing to the anthropic principle to fix the observed cosmological constant, it is not a large step to imagine the possibility that other hierarchies may also exist, as those which enter in split-SUSY theories. In these theories there is a badly broken supersymmetry in which the masses of squarks, sleptons and other scalars are very high, of $ O(M_X)$, but the gauginos, as well as one scalar (which plays the role of the SM Higgs boson), have masses of $O(M_W)$. Obviously, in these theories the hierarchy problem ( why $M_H<< M_X$) remains, but compared to the fine tuning associated with the cosmological constant, this is a minor sin!

As Vempati discussed, \cite{Vempati} these models are relatively easy to construct and they retain the best aspects of low energy SUSY theories, namely :

        i.            Unification of couplings (which are mostly influenced by having relatively light  gauginos) still occurs

      ii.            Neutralinos emerge  as dark matter

Because in these models, effectively, scalars are heavy, the flavor problem of SUSY theories (a problem we will discuss below) is eliminated. The landscape ideas, as exemplified by these split-SUSY theories, typify well a shift in perspective of theorists about low energy supersymmetry. It appears now that simple extensions of the SM, like the MSSM, are less likely to be the way in which SUSY is realized in nature.

\section{Flavor Physics}

 \subsection{SUSY- insights and tribulations}

A remarkable aspect of the SM is that a number of processes which are not observed in nature, like flavor-changing neutral currents [FCNC], lepton-flavor violation [LFV], or electric dipole moments [edms], are automatically very suppressed. This is not the case in extensions of the SM. This was nicely illustrated at Moriond by Abel \cite{Abel} in his talk on intersecting d-brane models. These models produce too large a $\Delta M_K$ unless the string scale $M_s >> 10^7$ GeV.

Quite similar considerations apply, in general, to low energy supersymmetric extensions of the SM, where flavor violation is both a problem and an exciting window of discovery. LFV provides a nice example. Even starting with universal scalar SUSY breaking masses $m_{ij}= m_o\delta_{ij}$ at a high scale, renormalization group evolution produces non-diagonal masses at the weak scale, which serve to induce LFV. As Turzynski \cite{Tur} and Takanishi \cite{Takanishi} discussed, predictions for processes like  $\mu \to e \gamma$ are sensitive to the mass of heavy neutrinos. In particular, one needs $M_1> 10^{11}$  GeV if one wants to have a big enough branching ratio for this process [BR $> 10^{-14}$], so as to be accessible to future experiments. However, such large values for $M_1$ exacerbate the gravitino problem- a problem we already mentioned in the context of Leptogenesis.

 As was discussed by Lebedev \cite{Lebedev} and Farzan \cite{Farzan}, electric dipole moments provide another relevant example of constraints in SUSY models, related here to CP-violation. In the simplest case of flavor blind supergravity models, CP-violating phases can enter in four quantities: the gluino masses [$m_{1/2}$], the coefficient of the scalar Yukawa interactions [A], the Higgsino mass term [$\mu$], and in the bilinear scalar terms[$ B\mu]$. However, only two of these phases are physical, say those in A and B, since the other two can be rotated away. Because the experimental bounds on edms are so strong, this severely restrict these phases. Typically, \cite {Lebedev} one obtains $\sin \phi_{A,B} < 10^{-2}-10^{-3}$.

   In general, flavor violating contributions due to SUSY matter entering in loops need to be controlled and different ideas have been put forth for how to do this. Three broad ways have been suggested to keep these SUSY induced flavor breaking problems below present experimental limits: \cite{Nir} \newline
 i) {\it Universality}, where one controls the flavor splitting among spartners [$\Delta \tilde{m}^2<<\tilde{m}^2$]; \newline
 ii) {\it Alignment}, where one tries to make the effective low energy gluino couplings very close to diagonal [$\gamma_{\tilde{g}ij}\sim \delta_{ij}]$; \newline
   iii) By introducing a {\it gap} between the Fermi scale and the scale of superpartners, discussed by Lavignac \cite{Lavignac} in Moriond [$ \tilde{m} >> $TeV]. \newline
\noindent
If SUSY is found, it will be great fun to sort out how the flavor problem is really solved in these theories!

\subsection{News from the Kaon sector}

Two nice pieces of news concerning the strange quark sector were reported at Moriond:

        i.            New experimental results from KLOE \cite{Lanfranchi} and KTEV \cite{Bellantoni} allowed one to infer a new precise value for $V_{us}$ different from the one now quoted in the PDG, which restored the unitarity relation $ |V_{ud}|^2 + |V_{us}|^2 + |V_{ub}|^2 =1$.

  ii.            Substantial progress has been made in understanding long distance effects in the decay $K_L \to \pi \ell^+ \ell^-$. \cite{Smith} This came both as a result of new experimental information on the process $K_S \to \pi \ell^+ \ell^-$ and as a result of theoretical calculations in chiral perturbation theory, which established  that there is a {\bf{positive}} interference between the short distance and the long distance pieces of  the parity violating part of the $K_L$ amplitude for this process. \cite{Greynat}

\subsection {B-physics and the CKM model}

Morandin \cite{Morandin} reviewed the very
precise value for the CP-violating quantity $ \sin 2\beta$ obtained by Babar and Belle by studying a variety of B-decay channels. This value, 
\begin{equation}
\sin 2\beta= 0.726 \pm 0.037,
\end{equation}
as discussed by Bosch, \cite{Bosch} provides strong confirmation of the validity of the CKM model. This is shown very clearly by the small
size of the overlap region in the global fit of the CKM model \cite{CKM} in the $\rho-\eta$ plane shown in Fig. 1. 

\begin{figure}\begin{center}
\psfig{file=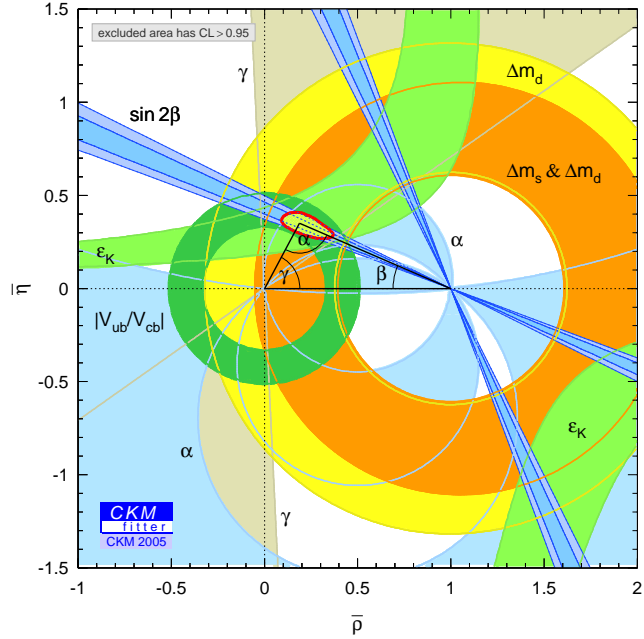,height=8.5cm,width=8.5cm}
\caption{Global fit of the CKM model in the $\rho-\eta$ plane.}
\end{center}
\end{figure}

As a result of the ongoing collaboration between experimentalists and theorists the extraction of the sides and (other) angles of the Unitarity Triangle in Fig. 1 is in the process of continuously being improved. A nice example of this was provided by the determination of $V_{ub}$ discussed by Limosani \cite{Limosani} at this meeting. As the result of more refined theoretical considerations, the theoretical uncertainties in $V_{ub}$ have now been reduced to 2.8\% for the shape function and 3.9\% for the value of $ m_b$, allowing for a 10\% determination of this quantity.

I was also very impressed by the great progress being made by Babar and Belle in extracting the other two angles in the Unitarity Triangle using an array of clever techniques (invented by theorists!). These results and the techniques used include:

     i.            Performing a Dalitz interference analysis of the $ B\to DK$ processes to extract the angle $\gamma$, yielding : \cite{Krokovney} 
\begin{equation} 
\gamma = (70 \pm 26 \pm 10 \pm 10)^{o}~~[{\rm Babar}] ;~~
\gamma = (68 \pm 14 \pm 13 \pm 11)^{o}~~ [{\rm Belle}] 
\end{equation}

ii.            The isospin analysis of the $ B \to \rho \rho$ processes  discussed by Wilson, \cite{Wilson} which produced the value for $\alpha$:
\begin{equation}
\alpha= (103 \pm 10)^{o}~~ [{\rm Babar}].
\end{equation}

   iii.            The mixing-decay interference analysis in the decays $ B \to D^*\rho$ used by Therin \cite{Therin} to extract, from combined data of Babar and Belle, the 68 \% CL bound 
\begin{equation}|
\sin(2\beta +\gamma)| >0.74. 
\end{equation}

      \subsection{Peephole to new physics?}

All of the above results are consistent with the CKM model fit. However, with more data, and further analysis, chinks in the CKM armor may well appear. One such hint has surfaced already, but it is to early to tell if it does, or does not, signal the presence of new physics.

Morandin \cite{Morandin} discussed an apparent discrepancy seen in the value of $\sin 2\beta$ obtained by Babar and Belle for B-decays, like $B\to \phi K_S$, which are dominated by Penguin modes. For these modes, rather that obtaining the value given in Eq. (17), one finds
\begin{equation} 
               \sin2\beta_{\rm{eff}}=0.43 \pm 0.07.
\end{equation}
Because the results of the two experiments for $\sin2\beta_{\rm {eff}}$ for the Penguin modes is systematically lower than that of the CKM fit, in principle one can imagine that new physics, like SUSY, could modify the Penguin contribution, thereby causing $\beta_{\rm {eff }} \neq \beta$.                                             This may indeed be the case, since the SM Penguin graph has no weak phase in the dominant loops, so even small "new Physics" effects could make a big difference.

However, in my view, before making too strong claims one must check that there are not more mundane answers (like QCD, or final state corrections) which could affect Penguin B-decay modes. Furthermore, one also needs to check that whatever "new physics" effects one adduces to explain why $\beta_{\rm{ eff }} \neq \beta$ does not affect other processes. In particular, the fact that the process $ B \to s\gamma$ seems to agree with the SM expectations puts strong constraints on what may be allowed. Obviously, time will tell!

\section*{Acknowledgements}
This work was supported in part by the Department of Energy under Contract No. FG03-91ER40662, Task C. I am very grateful to J. Tran Thanh Van, J. -M. Frere and F. Montanet for their hospitality in La Thuile.

 \end{document}